# Molecular Dynamics Simulation Study of Carbon-Nanotube Oscillator in Graphene Nanoribbon Trench

Eunae Lee[1], Jeong Won Kang[1,2,]*, Ki-Sub Kim[3,]**, Oh-Kuen Kwon[4,]***

[1] Department of IT Convergence, Korea National University of Transportation, Chungju 380-702, Republic of Korea

[2] Graduate School of Transportation, Korea National University of Transportation, Uiwang-si, Gyeonggi-do 437-763, Republic of Korea and

[3] Department of Chemical and Biological Engineering, Korea National University of Transportation, Chungju 380-702, Republic of Korea

[4] Department of Electronic Engineering, Semyung University, Jecheon 390-711, Republic of Korea

*Corresponding Author.

Tel.: +82-31-462-8739; Fax: +82-31-462-8734.

E-mail address: jwkang@ut.ac.kr

**Corresponding Author.

E-mail address : kks1114@ut.ac.kr

***Corresponding Author.

E-mail address : kok1027@semyung.ac.kr

ABSTRACT

Graphene/carbon-nanotube (CNT) hybrid material can be useful in energy storage and nanoelectronic technologies. Here we address the CNT-oscillator encapsulated in a graphene-nanoribbon (GNR) trench as a novel design, and investigate its properties via classical molecular dynamics simulations. Since the energy barrier was very low while the CNT was encapsulated in the GNR trench, the CNT absorbed on the GNR surface could easily be encapsulated in the GNR trench. MD simulations showed that the CNT oscillator encapsulated in a GNR trench is compatible with simple CNT oscillators, so we anticipate that the CNT in GNR trench could work as an oscillator. So we can anticipate that the CNT encapsulated in a GNR trench can be applied to ultra-sensitive nanoelectromechanical oscillators, and this system has the possibility to be applied to relay-switching devices, and to shuttle memory.

# 1. INTRODUCTION

Graphene [1], a one-atom thick layer of graphite, has recently received a lot of attention, due to its impressive mechanical and electromagnetic properties [2-4], including zero electron bandgap, high electron emission rate, and elastic scattering [5-7]. Carbon nanotube (CNT) [8], a quasi-one-dimensional nanostructure, can be thought of as graphene rolled up into a cylinder. Both CNTs and graphenes have a unique place in nanoscience, owing to their exceptional electrical, thermal, chemical, and mechanical properties; and the number of their potential applications continues to grow [9].

As one of the engineering applications based on CNTs, gigahertz CNT oscillators were proposed by Zheng and Jiang [10], after Cumings and Zettl [11] reported an ideal low-friction and low-wear bearing carved out of a multi-walled CNT with a diameter of a few tens of nanometers, and they have been intensively investigated [12-19]. Moreover, graphene oscillators based on the interlayer sliding of graphene nanoribbons (GNRs) or graphene nanoflakes (GNFs) have been addressed [20-28].

Recently, conjugated carbon nanomaterials, such as fullerene-nanotube, fullerene-graphene, and nanotube-graphene hybrids, also have great potential application [29,30]. Fullerene-nanotube hybrids termed carbon nanopeapods have been widely investigated [31-33]. Quite recently, fullerene-graphene hybrids have been experimentally produced and theoretically investigated [34-39]. In particular, nanotube-graphene hybrid also has great potential application [40-48]. Theoretical work suggested that a covalently bonded graphene-nanotube hybrid material would extend those properties to three dimensions, and be useful in energy storage and nanoelectronic technologies [40]. Zhu et al. [40] disclosed a method of seamlessly bonding graphene and single-walled CNTs, during the growth stage. The hybrid material exhibited a surface area > 2,000 $m^2 g^{-1}$, with ohmic contact from the vertically aligned single-walled CNTs to the graphene. Using aberration-corrected scanning transmission electron microscopy, they observed the covalent transformation of $sp^2$ carbon between the planar graphene and the single-walled CNTs, at the atomic resolution level. Their findings provided a new benchmark for understanding three-dimensional graphene-nanotube-conjoined materials.

In our previous work [49], using CNT-GNR hybrid, we proposed nonvolatile CNT-shuttle-memory on GNR array, and investigated its dynamics via classical molecular dynamics (MD) simulations. In this work, we present CNT-oscillator encapsulated in GNR trench as a CNT-GNR hybrid, and its operational properties, using atomistic simulations. The simulation methods and structures are addressed in Sections 2 and 3, respectively; and the simulation results are analyzed and discussed in Section 4. We conclude this paper in Section 5.

## 2. Simulation Methods

We used two empirical potential functions to study the CNT-oscillator encapsulated in a GRN trench. Interactions between carbon atoms that form the covalent bonds of GNR and CNT were described using the Tersoff-Brenner potential [50, 51], which has been widely applied in carbon systems. The long-range interactions of carbons between GNR and CNT were characterized according to the Lennard-Jones 12–6 (LJ12–6) potential, with the parameters that were given by Mao et al. [52]. In this paper, the parameters of the LJ12–6 potential were $\varepsilon_{carbon}$ = 0.0042 eV and $\sigma_{carbon}$ = 3.37 Å. The cutoff distance of the LJ12-6 potential was 10 Å.

To obtain the static properties, we used an optimization scheme, namely the steepest descent (SD) method [53], which is the simplest of the gradient methods, from the atomic configurations, with a C–C bond length of 0.142 nm. To characterize the dynamics properties, we used the same MD methods as those used in our previous studies [54-59]. The MD code uses the velocity Verlet algorithm and neighbor lists, to improve the computing performance. The MD time step ($\Delta t$) was $5 \times 10^{-4}$ ps. The temperatures were set at 1 K for all MD simulations, and the total energy was constant, during all the MD simulations.

# 3. Simulation Structure

Recently, substrate-regulated graphene morphologies have been intensively investigated [60, 61]. In particular, Hicks et al. [60] presented GNR trenches that enable a substantial electronic bandgap, by coating bi-layer graphene on silicon carbide nanometer-scale steps. GNR trench morphology can be formed on the trench of the substrate surface, through the adhesion of graphene on the substrate, due to their attractive forces [60]. Such a GNR trench possesses large empty spaces inside of the trench, which are like the large empty spaces inside carbon nanotubes (CNTs) [31]. So the 'carbon nanopeapod (CNP)' has been reported, consisting of fullerene arrays inside single-walled CNTs, [32]. Therefore, GNR trench morphology with empty spaces can also open new applications for graphenes, as storage materials with high capacity and high stability [62]. In our previous work [62], two mechanisms of $C_{60}$ encapsulation into the GNR trench were investigated, and the probability of a nanocontainer based on GRN trench in the nanometer ranges was suggested. Surely, CNTs can also be encapsulated in the GNR trench. Like the $C_{60}$-CNT oscillators, which have been investigated [63-65], the CNT encapsulated in a GNR trench has an engineering application, as an oscillator based on CNT-GNR hybrid, as shown in Figure 1.

Hence, to study the CNT-oscillator encapsulated in a GRN trench, we considered a model structure, as shown in Fig. 1. The capped (10,10) CNTs were each composed of 250 carbon atoms, with the length of 2.9 nm ($L_{CNT}$). Zigzag-edge graphene with a width of 2 nm and length of 10 nm was composed of 570 carbon atoms, and the trench was of 2 nm width and 2 nm depth. First, the GNR was initially relaxed, without the CNT. Afterwards, the CNT was placed on top of the GNR trench; then the encapsulation processes were characterized by both the SD and the MD methods; and finally, the oscillation features were analyzed, by the data obtained from the MD simulations.

# 4. RESULTS AND DISCUSSION

## 4.1 CNT Encapsulation into GNR Trench

Figure 1 shows the schematics of the CNT oscillator based on a CNT-GNR hybrid. We performed MD simulations to obtain the CNT oscillator encapsulated in a GNR trench, when the CNT was mounted on the GNR trench. Figure 2 shows the variations of the van der Waals (vdW) energy ($U_{vdW}$) between the CNT and the GNR, as a function of the MD time. The CNT was directly sucked into the GNR trench, because of the attractive forces between the CNT and GNR, except that the CNT experienced little interference at the entrance of the trench.

In the case of Figure 2, since the CNT was located in the center of the GNR trench, the encapsulating dynamics were achieved without any energy barrier. However, in general, the CNT would be randomly deposited in the GNR trench, and thus the situation will require some energy barriers for the encapsulation dynamics. For the fullerene-CNT hybrids, Ulbricht and Hertel [66] discussed the encapsulation via different reaction paths through open tube ends or sidewall defects, and compared the capture of the $C_{60}$ molecule from the gas phase, with the encapsulation of $C_{60}$ preadsorbed on the surface of the CNT rope. They obtained suction energy barriers of 0.31~ 0.55 eV for the (10, 10) CNT. However, some previous works have shown no energy barrier of $C_{60}$ encapsulations into CNTs [67-70]. Berber et al. [67], using tight-binding MD simulations, also showed that the encapsulation process did not involve an activation energy barrier. Qian et al. [70] showed that the $C_{60}$ was sucked into the (10, 10) CNT, by the sharp surface tension force presented in the front of the open end.

To obtain the energetics of the CNT-oscillator encapsulated in the GNR trench, first the CNT was mounted on the GNR surface, and then the CNT-GNR hybrid was relaxed by the SD method. After the relaxation, the CNT was slightly displaced; and then the CNT-GNR hybrid was once more relaxed by the SD method. This process was repeated, until the CNT was fully encapsulated in the GNR trench. Figure 3 shows the variations of the $U_{vdW}$, against the position of the CNT axis. In Figure 3, we can find the energy barrier of ~ 0.52 eV, during the CNT being encapsulated in the GNR trench. This value is higher than the encapsulation energy barrier of ~0.2 eV, when the $C_{60}$ molecule is encapsulated in a GNR trench [62].

Since Figure 3 shows the results in the static condition via the SD method, we performed MD simulations, to present the dynamic results during the encapsulation. Figure 4 (a) shows the variation of the $U_{vdW}$ as a function of the MD time, when the CNT was deposited on the GNR surface (Case 1). In this work, the binding energy of the CNT on the GNR surface was ~1.8 eV. For this case, the energy barrier of ~ 0.5 eV was found, when the CNT reached the edge of the GNR trench. This value is in good agreement with the 0.52 eV found in Figure 3. In particular, the energy barrier in this work is 2.08 meV/atom, which corresponds to a temperature of 16 K. So we can discuss that the CNT can be easily sucked into the GNR trench by just its thermal energy, without any externally applied force. Figure 4 (b) shows the variation of the $U_{vdW}$ as a function of the MD time, when the CNT was deposited on the edge of the GNR trench (Case 2). This case did not show any energy barrier. Figure 4 (c) shows the variations of the $U_{vdW}$ as a function of the position of the CNT ($C_x$). Clearly, these results show that the GNR trench provides the energetically stable space for the CNT absorbed on the GNR surface.

### 4.2 Oscillatory Behaviors of CNT-Oscillator

To study the oscillatory behaviors of the CNT-oscillator in the GNR trench, first, we obtained the optimized position of the CNT in the GNR trench via the SD scheme, as shown in Figure 5 (a); and afterwards, the position of the CNT was displaced along the direction of the GNR trench, as shown in Figures 5 (b) and (c). Then, the vdW energy and force between the CNT and the GNR trench were calculated, as a function of the position of the CNT. Figure 6 present the vdW energy and force as a function of the displacement ($x$) of the CNT, respectively. When the CNT was displaced from the center of the GNR trench, the vdW energy ($U_{vdW}$) increased, with increasing displacement of the CNT, as shown in Figure 6. Such a change of the vdW energy against the displacement is related to the induced vdW force ($F_{vdW}$), $F_{vdW} = -dU_{vdW} / dx$. The induced vdW forces were always suction forces. When the CNT was extruded from the GNR trench, as shown in Figures 5 (b) and (c), the restoring forces of the CNT were always toward the center of the GNR trench, as shown in the vdW force plot in Figure 6. So these vdW forces can be approximated by two equal and opposite Dirac delta functions, operating at both ends of the GNR trench [71]. Hence, the extruded CNT can quickly and fully retract inside the GNR trench, owing to the restoring force resulting from the vdW interaction acting on the extruded CNT, and then oscillate between the two open ends of the GNR trench. As a result, high frequency can be generated by a CNT oscillating inside a GNR trench.

Next, we performed MD simulations for the CNT with different initial velocities ($v_0$). In general, the critical parameters for the CNT oscillators are its initial velocity and displacement, the magnitude and frequency of the driving force, and their size. The oscillatory behavior of the CNT oscillators in the GNR trench can be initiated by precision mechanical controllers; and it can be remotely initiated by electromagnetic fields, lasers and electrostatic forces [12]. The resonance frequencies of the CNT oscillators in the GNR trench are dependent on the length and width of the GNR trench, as well as the initial velocity; and hence, their operating frequencies can be adjusted, by controlling their structural parameters and initial conditions.

In this work, we assumed that the CNT had a velocity, due to the externally applied impulse force. Figure 7 shows the displacements, the vdW energies, and the vdW forces of the CNT inside the GNR trench, as a function of the MD time, for $v_0 = 0.65$ nm/ps. Liu et al. [72] investigated the oscillatory behavior of a $C_{60}$ oscillating through an SWCNT, and found that the $C_{60}$ performs a decaying oscillation; and that the decaying of the oscillatory amplitude and the oscillatory frequency are sensitive to the diameter and degree of helicity of the nanotube. It can be seen from Figure 7 that the amplitude of the oscillation gradually decreases with oscillatory time. In this work, when $v_0 > 0.65$ nm/ps, the CNT escaped from the GNR trench.

Surely, the translation motion of the CNT along the GNR trench is dominant. During the oscillation, the decrease of the oscillatory amplitude indicates that portions of the potential energy and kinetic energy of the CNT, although small in each cycle of motion, inevitably convert into random thermal vibration energy, and lead to the energy dissipation of the nano-oscillator [73, 74]. It is found that the CNT performed a decaying oscillation, with a gradual decrease of the oscillatory amplitude, and a gradual decrease of the oscillatory frequency. It is shown that this decay is due to the transformation of thermal energy [72]. Actually, the CNT oscillator in a GNR trench, with thousands of degrees of freedom, can be reduced to a simple system with a few most relevant degrees of freedom, in the presence of a thermal bath. Zhao et al. [73] discussed the energy exchange mechanisms for the CNT oscillator. A few degrees of freedom correspond to several important low-frequency mechanical modes, such as axial translational motion, orthogonal vibrational motion, and rotational motion; while the heat bath is made of other higher-frequency vibrations of the GNR. When the kinetic energy leakage from the reduced system to the bath is slow enough, the energy exchange takes place between the axial translational motion, and the orthogonal vibrational and rotational motions. The three quasi-stable regimes are linked to form a reduced system, which allows energy exchanges among them, but otherwise prohibits a rapid thermal equilibration from taking place [73]. An interesting feature of this reduced system is that the energy exchanges among quasi-stable modes are statistical in nature, unlike those in an isolated mechanical system with a few degrees of freedom [73]. The axial oscillation of the CNT oscillator in the GNR trench is the mechanical mode of interest in NEMSs, decreasing in both frequency and amplitude by the dissipative heat bath.

Figure 8 (a) shows the displacements of the CNT inside the GNR trench, as a function of the MD

time, for four $v_0$ values. The displacements of the CNT increase with increasing $v_0$. Figure 8 (b) shows the amplitude spectra as a function of the frequency ( $f$ ), for four $v_0$ values. These spectra explicitly present the resonance frequencies with dominant peaks. Figure 8 (c) shows the resonance frequency as a function of the $v_0$. The frequency increases until $v_0 = 0.2$ nm/ps, and then gradually decreases with increasing $v_0$. For the simple CNT oscillators, Legoas et al. [14, 15] suggested that their frequencies can be estimated as follows

$$f \approx \frac{1}{2} \left( \frac{2 m_{CNT} v_0}{F_{Max}} + \frac{\Delta L}{v_0} \right)^{-1} , \qquad (1)$$

where, $m_{CNT}$ and $\Delta L$ are the mass of the CNT, and the length difference between the width of the GNR trench and the length of the CNT, respectively. $F_{Max}$ is the maximum value of the $F_{vdW}$. The results obtained from the MD simulations in this work are compared to the data obtained from Eq. (1), as shown in Figure 8 (c). Here $\Delta L = 0.9$ nm, $m_{CNT} = N_C \times 1.99 \times 10^{-26}$ kg $= 497.5 \times 10^{-26}$ kg, and $F_{Max} = 0.5$ nN. When $v_0 \geq 0.15$ nm/ps, the MD results are in good agreement with the estimation using Eq. (1). Therefore, the dynamics motions of the CNT oscillator in the GNR trench in this work are similar to those for the simple CNT oscillators. The trend of the $f$-$v_0$ plot in this work is similar to some cases of the previous work [74], which investigated the dynamics properties of the CNT oscillators with intertube gaps, via classical MD simulations.

Although the system considered in this work was very difficult to be realized in the general experiments, our results explicitly addressed the potential application of the CNT encapsulated in the GNR trench to ultra-sensitive nanoelectromechanical oscillators. However, we should note that, although the CNT oscillators have been extensively investigated in theoretical works, their operational frequencies obtained from experiments have never been presented. There are many technological challenges that currently hinder the realization and implementation of the proposed CNT oscillators. Fabricating CNT oscillators represents tough technological challenges, because it requires handling parts that are nanoscale in size, and parts that oscillate in the gigahertz range. Furthermore, the difficulties in packing such CNT oscillators, and realizing the communicating signals from the nanoscale to the macroscopic world, also hinder the implementation of the proposed oscillators. However, although the current CNT oscillator in GNR trench is difficult to implement by simple processes, we should note that such a structure can be achieved by the nanotube-graphene hybrid nanofabrication processes, as discussed in the Introduction [40-48].

This result shows the possibility that the CNT-oscillators in the GNR trench can be applied to a relay-switching device, or a shuttle memory. Therefore, further work should include quantum mechanical properties, to reveal the structural and electronic properties of unusual carbon nanostructures, electron transport properties, and the influence of electric and magnetic fields [76-86]. The mechanical quality factor also plays a very important role in determining the sensitivity of nanoelectromechanical-based devices. One can expect that by increasing the temperature, the quality factor will be greatly decreased [75]. Therefore, further work should include MD simulations carried out at a substantially higher temperature range than the 1 K value in this work, to consider thermal dissipation, because real experiments are most likely to be conducted at much higher temperatures than 1 K.

## IV. CONCLUDING REMARKS

The CNT-oscillator encapsulated in a GNR trench was addressed, and its properties were investigated via classical MD simulations. Since the energy barrier while the CNT was encapsulated in the GNR trench was very low at 2.08 meV/atom, the CNT absorbed on the GNR surface could easily be encapsulated in the GNR trench. The motions of the CNT in the GNR trench were similar to those of the core CNT encapsulated in an outer CNT, and as a result, the CNT in a GNR trench can work as an oscillator. The results obtained from the MD simulations in this work showed that the CNT oscillator encapsulated in a GNR trench is compatible with simple CNT oscillators. So we can note that the CNT encapsulated in a GNR trench can be applied to ultra-sensitive nanoelectromechanical oscillators; and this system has the possibility of being applied to a relay-switching device, and to shuttle memory. Finally, the authors anticipate the realization of this proposed system. In order that the proposed system will operate effectively, the oscillation frequency, the applied force field, the active region, the length of the CNT, and the size of GNR trench should be considered in future work.

## ACKNOWLEDGEMENT

This research was partially supported by the MSIP (Ministry of Science, ICT and Future Planning), Korea, under the C-ITRC (Convergence Information Technology Research Center) (IITP-2015-H8601-15-1008) supervised by the IITP (Institute for Information & communications Technology Promotion), and partly supported by the Ministry of Education (MOE) and National Research Foundation of Korea (NRF) through the Human Resource Training Project for Regional Innovation (2014H1C1A1066414).

**Figure captions**

Figure 1. Model schematics of the CNT oscillator based on a CNT-GNR hybrid. The capped (10,10) CNT was composed of 250 carbon atoms, with the length of 2.9 nm ($L_{CNT}$). Zigzag-edge graphene with a width of 2 nm and length of 10 nm was composed of 570 carbon atoms, and the trench was 2 nm width and of 2 nm depth.

Figure 2. The variations of the van der Waals energy ($U_{vdW}$) between the CNT and the GNR as a function of the MD time are shown, of when the CNT was directly sucked into the GNR trench.

Figure 3. Variations of the $U_{vdW}$ against the position of the CNT axis. The energy barrier of ~ 0.52 eV was found while CNT was encapsulated in the GNR trench.

Figure 4. MD simulations results. (a) Variation of the $U_{vdW}$ as a function of the MD time, when the CNT was deposited on the GNR surface (Case 1). (b) Variation of the $U_{vdW}$ as a function of the MD time, when the CNT was deposited on the edge of the GNR trench (Case 2). (c) Variations of the $U_{vdW}$ as a function of the position of the CNT ($C_x$).

Figure 5. (a) Optimized position of the CNT in the GNR trench. The displacements of the CNT toward the (b) left, or the (c) right sides, are shown.

Figure 6. Variations of the vdW energy ($U_{vdW}$) and the vdW force ($F_{vdW}$), as a function of the position of the CNT.

Figure 7. MD simulation results for $v_0 = 0.65$ nm/ps. Variations of (a) displacements, (b) $U_{vdW}$, and (c) $F_{vdW}$, as a function of the MD time.

Figure 8. (a) Displacements of the CNT inside the GNR trench, as a function of the MD time, for four $v_0$ values. (b) Amplitude spectra, as a function of the frequency ($f$), for four $v_0$ values. (c) Resonance frequency, as a function of the $v_0$.

# References

[1] K. S. Novoselov, A. K. Geim, S. V. Morozov, D. Jiang, Y. Zhang, S. V. Dubonos, A. A. Firsov, *Science* 306, 666 (2004).

[2] A. K. Geim, *Science* 324, 1530 (2009).

[3] A. K. Geim, K.S. Novoselov, *Nature Mater.* 6, 183 (2007).

[4] A. H. Castro Neto, F. Guinea, N. M. R. Peres, K. S. Novoselov, A. K. Geim, *Rev. Mod. Phys.* 81, 109 (2009).

[5] K. S. Novoselov, A. K. Geim, S. V. Morozov, D. Jiang, M. I. Katsnelson, I. V. Grigorieva, S. V. Dubonos, A. A. Firsov, *Nature* 438, 197 (2005).

[6] Y. Zhang, Y. W. Tan, H. L. Stormer, P. Kim, *Nature* 438, 201 (2005).

[7] K. S. Novoselov, D. Jiang, F. Schedin, T. J. Booth, V. V. Khotkevich, S. V. Morozov, A. K. Geim, *Proc. Natl. Acad. Sci. USA* 102, 10451 (2005).

[8] S. Iijima, *Nature* 354, 56 (1991).

[9] T. C. Dinadayalane, Jerzy Leszczynski, *Structural Chemistry* 21, 1155 (2010).

[10] Q. Zheng, Q. Jiang, *Phys. Rev. Lett.* 88, 045503 (2002).

[11] J. Cumings, A. Zettl, *Science* 289, 602 (2000).

[12] J. W. Kang, H. J. Hwang, *J. Comput. Theor. Nanosci.* 6, 2347 (2009).

[13] J. Cumings, Z. Zettl, *Phys. Rev. Lett.* 93, 086801 (2004).

[14] S. B. Legoas, V. R. Coluci, S. F. Braga, P. Z. Coura, S. O. Dantas, D. S. Galvão, *Phys. Rev. Lett.* 90, 055504 (2003).

[15] S. B. Legoas, V. R. Coluci, S. F. Braga, P. Z. Coura, S. O. Dantas, D. S. Galvão, *Nanotechnology* 15, S184 (2004).

[16] J. W. Kang, H. J. Hwang, *J. Appl. Phys.* 96, 3900 (2004).

[17] J. W. Kang, K. O. Song, O. K. Kwon, H. J. Hwang, *Nanotechnology* 16, 2670 (2005).

[18] J. W. Kang, K. O. Song, H. J. Hwang, Q. Jiang, *Nanotechnology* 17, 2250 (2006).

[19] A. Neild, T. W. Ng, A. Zheng, *Euro. Phys. Lett.* 87, 16002 (2009).

[20] Q. Zheng, B. Jiang, S. Liu, Y. Weng, L. Lu, Q. Xue, J. Zhu, Q. Jiang, S. Wang, L. Peng, *Phys. Rev. Lett.* 100, 067205 (2008).

[21] I. V. Lebedeva, A. A. Knizhnik, A. M. Popov, Yu. E. Lozovik, B. V. Potapkin, *Phys. Chem. Chem. Phys.* 13, 5687 (2011).

[22] I. V. Lebedeva, A. A. Knizhnik, A. M. Popov, Yu. E. Lozovik, B. V. Potapkin, *Physica E* 44, 949 (2012).

[23] A. M. Popov, I. V. Lebedeva, A. A. Knizhnik, Y. E. Lozovik, and B. V. Potapkin, *Phys. Rev. B* 84, 245437 (2011).

[24] Z. Liu, P. Boggild, J.-R. Yang, Y. Cheng, F. Grey, Y.-L. Liu, L. Wang, Q.-S. Zheng, *Nanotechnology* 22, 265706 (2011).

[25] I. V. Lebedeva, A. A. Knizhnik, A. M. Popov, O. V. Ershova, Yu. E. Lozovik, B. V. Potapkin, *Phys. Rev. B* 82, 155460 (2010).

[26] I. V. Lebedeva, A. A. Knizhnik, A. M. Popov, O. V. Ershova, Yu. E. Lozovik, B.V. Potapkin, *J. Chem. Phys.* 134, 104505 (2011).

[27] A. M. Popov, I. V. Lebedeva, A. A. Knizhnik, Y. E. Lozovik, B. V. Potapkin, *Phys. Rev. B* 84, 245437 (2011).

[28] O. K. Kwon, H.-W. Kim, J. W. Kang, *Curr. Appl. Phys.* 14, 237 (2014).

[29] D. Huertas-Hernando, F. Guinea, A. Brataas, *Phys. Rev. B* 74, 155426 (2006).

[30] S.-Y. Kim, H. J. Hwang, J. W. Kang, *Phys. Lett. A* 377, 3136 (2013).

[31] B.W. Smith, M. Monthioux, D. E. Luzzi, *Nature* 396, 323 (1998).

[32] M. Monthioux, *Carbon* 40, 1809 (2002).

[33] S. Berber, Y. K. Kwon, D. Tománek, *Phys. Rev. Lett.* 88, 185502 (2002).

[34]    A. V. Savina, Y. S. Kivshar, *Scientific Reports* 2, 1012 (2012).
[35]    M. Ishikawa, S. Kamiya, S. Yoshimoto, M. Suzuki, D. Kuwahara, N. Sasaki, K. Miura, *J. Nanomater.* 2010, 891514 (2010).
[36]    M. Švec, P. Merino, Y. J. Dappe, C. González, E. Abad, P. Jelínek, J. A. Martín-Gago, *Phys. Rev. B* 86, 121407 (2012).
[37]    M. Neek-Amal, N. Abedpour, S. N. Rasuli, A. Naji, M. R. Ejtehadi, *Phys. Rev. E* 82, 051605 (2010).
[38]    N. Itamura, H. Asawa, K. Miura, N. Sasaki, *J. Phys.: Conference Series* 258, 012013 (2010).
[39]    A. Lohrasebi, M. Neek-Amal, M. R. Ejtehadi, *Phys. Rev. E* 83, 042601 (2011).
[40]    Y. Zhu, L. Li, C. Zhang, G. Casillas, Z. Sun, Z. Yan, G. Ruan, Z. Peng, A.-R. O. Raji, C. Kittrell, R. H. Hauge, J. M. Tour, *Nature Communications* 3, 1225 (2012).
[41]    Z. Fan, J. Yan , L. Zhi, Q. Zhang, T. Wei, J. Feng, M. Zhang, W. Qian, F. Wei, *Adv. Mater.* 22, 3723 (2010).
[42]    R. Lv, T. Cui, M.-S. Jun, Q. Zhang, A. Cao, D. S. Su, Z. Zhang, S.-H. Yoon, J. Miyawaki, I. Mochida, F. Kang, *Adv. Funct. Mater.* 21, 999 (2011).
[43]    Z. Sui, Q. Meng, X. Zhang, R. Ma, B. Cao, *J. Mater. Chem.* 22, 8767 (2012).
[44]    L. Peng, Y. Feng, P. Lv, D. Lei, Y. Shen, Y. Li, W. Feng, *J. Phys. Chem. C* 116, 4970 (2012).
[45]    X. Dong, G. Xing, M. B. Chan-Park, W. Shi, N. Xiao, J. Wang, Q. Yan, T. Chien Sum, W. Huang, P. Chen, *Carbon* 49, 5071 (2011).
[46]    X. Chen, J. Zhu, Q. Xi, W. Yang, *Sensors and Actuators B: Chemical* 161, 648 (2012).
[47]    Z. Yan, L. Ma, Y. Zhu, I. Lahiri, M. G. Hahm, Z. Liu, S. Yang, C. Xiang, W. Lu, Z. Peng, Z. Sun, C. Kittrell, J. Lou, W. Choi, P. M. Ajayan, J. M. Tour, *ACS Nano* 7, 58 (2013).
[48]    J. M. Tour, *Chem. Mater.* 26, 163 (2014).
[49]    J. W. Kang, Z. Hwang, *J. Comput. Theor. Nanosci.* 12, 425 (2015).
[50]    J. Tersoff, *Phys. Rev. B* 39, 5566 (1989).
[51]    D. W. Brenner, *Phys. Rev. B* 42, 9458 (1990).
[52]    Z. Mao, A. Garg, S. B. Sinnott, *Nanotechnology* 10, 273 (1999).
[53]    J. A. Snyman, *Practical Mathematical Optimization: An Introduction to Basic Optimization Theory and Classical and New Gradient-Based Algorithms*, Springer Publishing, New York (2005).
[54]    J. W. Kang, K.-S. Kim, H. J. Hwang, *Computat. Mater. Sci.* 50, 1818 (2011).
[55]    J. W. Kang, K.-S. Kim, J. Park, H. J. Hwang, *Physica E* 43, 909 (2011).
[56]    J. W. Kang, K.-S. Kim, H. J. Hwang, *Computat. Mater. Sci.* 51, 216 (2012).
[57]    O. K. Kwon, G.-Y. Lee, H. J. Hwang, J. W. Kang, *Physica E* 44, 194 (2012).
[58]    K.-S. Kim, H. J. Hwang, J. W. Kang, *Physica E* 44, 1543 (2012).
[59]    J. W. Kang, K.-S. Kim, K. R. Byun, H. J. Hwang, *Phys. Lett. A* 375, 1470 (2011).
[60]    J. Hicks, A. Tejeda, A. Taleb-Ibrahimi, M. S. Nevius, F. Wang, K. Shepperd, J. Palmer, F. Bertran, P. Le Fèvre, J. Kunc, W. A. de Heer, C. Berger, E. H. Conrad, *Nature Phys.* 9, 49 (2013).
[61]    S.-Y. Kim, H. J. Hwang, J. W. Kang, *Phys. Lett. A* 377, 3136 (2013).
[62]    S.-Y. Kim, J. W. Kang, *J. Comput. Theor. Nanosci.* 11, 2125 (October 2014).
[63]    P. Liu, Y. W. Zhang, C. Lu. *J. Appl. Phys.* 97, 094313 (2005).
[64]    H.-Y. Song, X.-W. Zha, *Phys. Lett. A* 373, 1058 (2009).
[65]    R. Ansari, F. Sadeghi, S. Ajori, *Mech. Res. Commun.* 47, 18 (2013).
[66]    H. Ulbricht, T. Hertel, *J. Phys. Chem. B* 107, 14185 (2003).
[67]    S. Berber, Y. K. Kwon, *Phys. Rev. Lett.* 88, 185502 (2002).
[68]    H. J. Hwang, K. R. Byun, J. W. Kang, *Physica E* 23, 208 (2004).
[69]    H. Ulbricht, G. Moos, T. Hertel, *Phys. Rev. Lett.* 90, 095501 (2003).
[70]    D. Qian, W. K Liu, R. S Ruoff, *J. Phys. Chem. B* 105, 10753 (2001).
[71]    J. W. Kang, K. W. Lee, *J. Korean Phys. Soc.* 65, 185 (2014).
[72]    L. Liu, Y. Su, J. Gao, J. Zhao, *Physica E* 46, 6 (2012).
[73]    N. Thamwattana, J. M. Hill, *J. Nanopart. Res.* 10, 665 (2008).

[74] Y. Zhao, C. C. Ma, G. H. Chen, Q. Jiang, *Phys. Rev. Lett.* 91, 175504 (2003).

[75] K.-S. Kim, J. Park, H. J. Hwang, J. W. Kang, *Physica E* 44, 2027 (2012).

[76] L. Wang, Q. Zhang, *Curr. Appl. Phys.* 12, 1173 (2012).

[77] S. Hwang, J. Lim, H.G. Park, W.K. Kim, D.H. Kim, I.S. Song, J.H. Kim, S. Lee, D.H. Woo, S.C. Jun, *Curr. Appl. Phys.* 12, 1017 (2012).

[78] M. Ashhadi, S.A. Ketabi, *Physica E* 46, 250 (2012).

[79] Y.-H. Zhang, L.-F. Han, Y.-H. Xiao, D.-Z. Jia, Z.-H. Guo, F. Li, *Comput. Mater. Sci.* 69, 222 (2013).

[80] K. P. Ghatak, S. Bhattacharya, A. Mondal, S. Debbarma, P. Ghorai, and A. Bhattacharjee, *Quantum Matter* 2, 25 (2013).

[81] I. S. Amiri, A. Nikoukar, J. Ali, and P. P. Yupapin, *Quantum Matter* 2, 42 (2013).

[82] A.F. Kuloglu, B. Sarikavak-Lisesivdin, S.B. Lisesivdin, E. Ozbay, *Comput. Mater. Sci.* 68, 18 (2013).

[83] P. Di Sia, *J. Comput. Theor. Nanosci.* 9, 31 (2012).

[84] X.-F. Peng, C. Xiong, X.-J. Wang, L.-Q. Chen, Y.-F. Luo, J.-B. Li, *Comput. Mater. Sci.* 77, 440 (2013).

[85] Z. Hwang, J.H. Lee, J.W. Kang, *J. Comput. Theor. Nanosci.* 10, 1892 (2013).

[86] N. Ding, X. Lu, C.M.L. Wu, *Comput. Mater. Sci.* 51, 141 (2012).

FIGURES

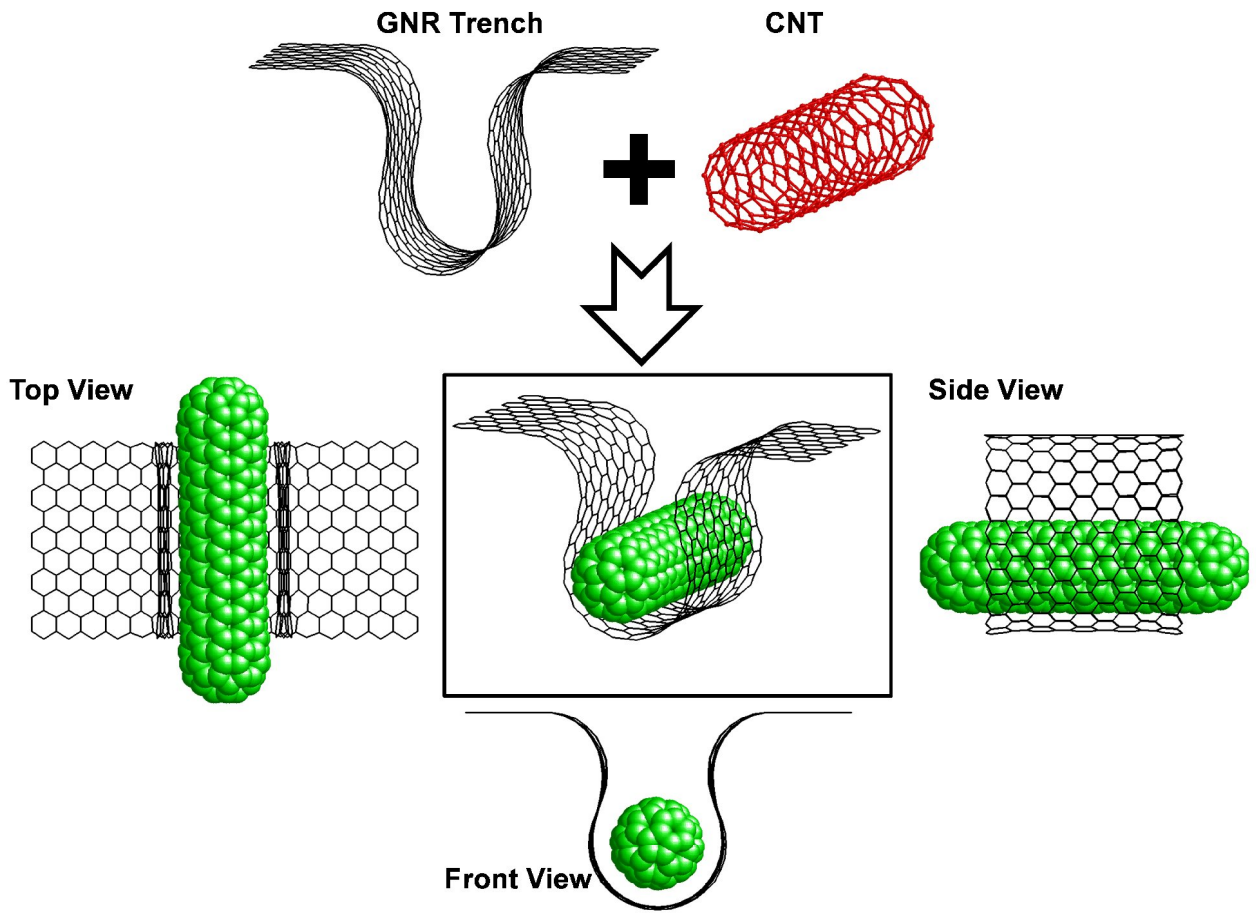

Figure 1.

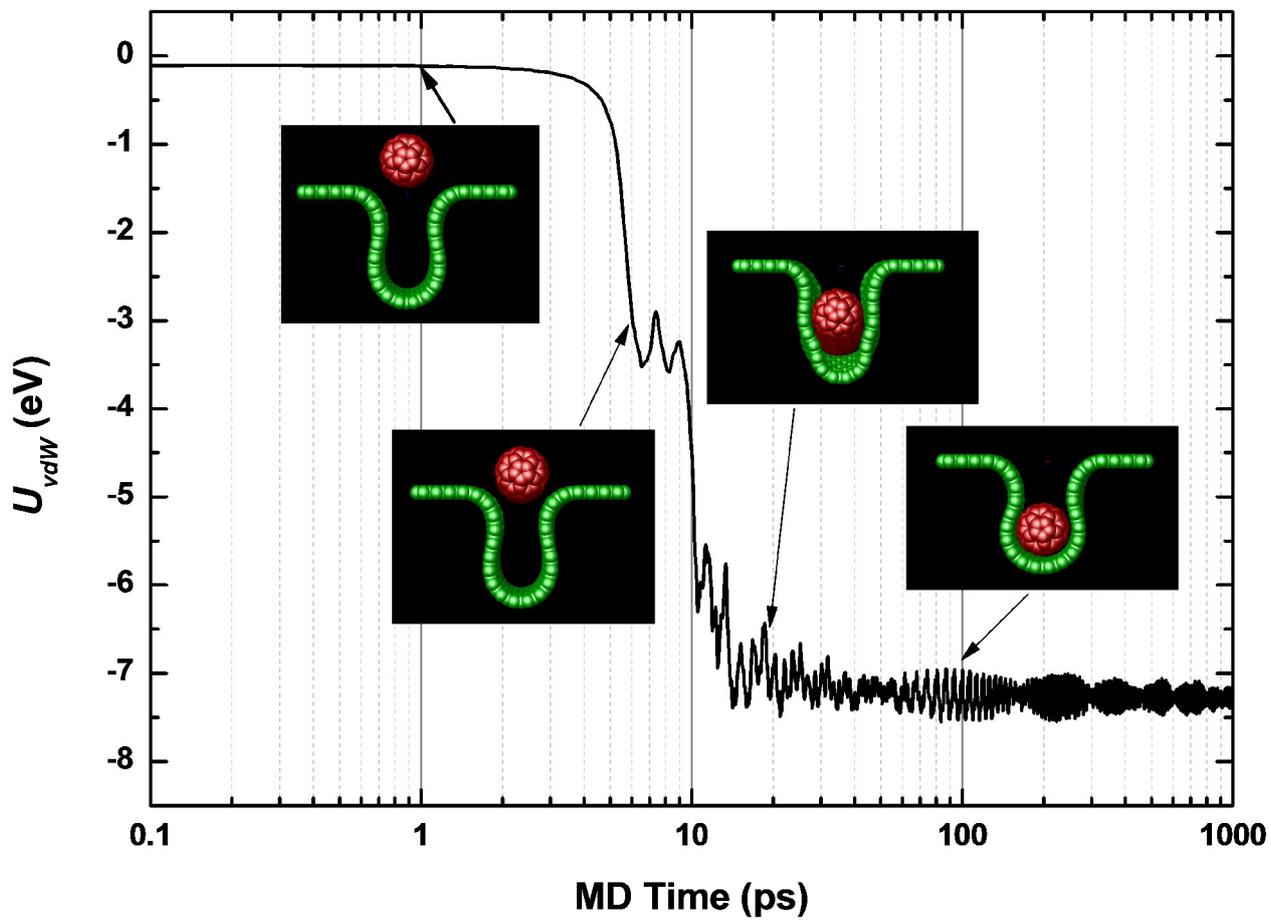

Figure 2.

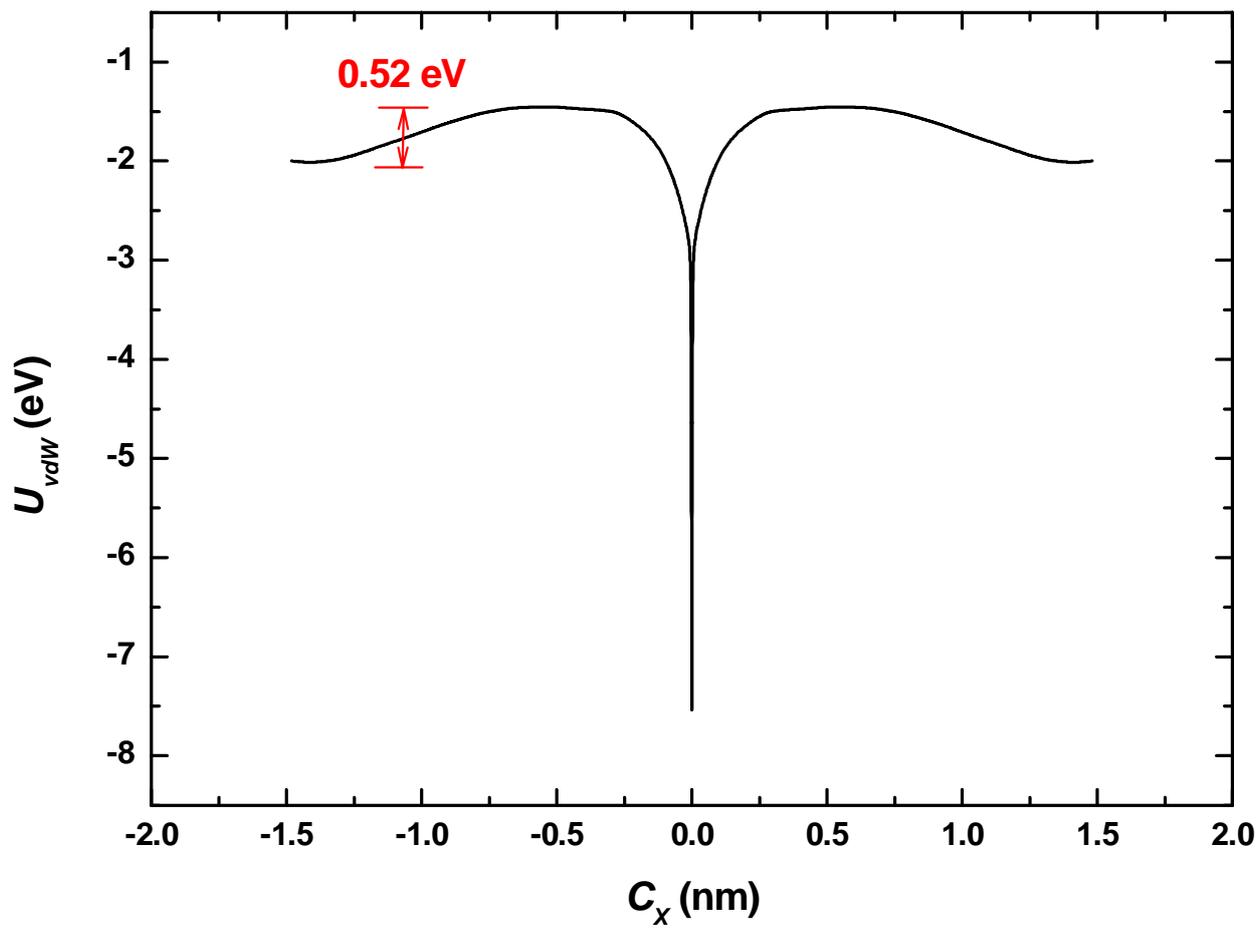

Figure 3.

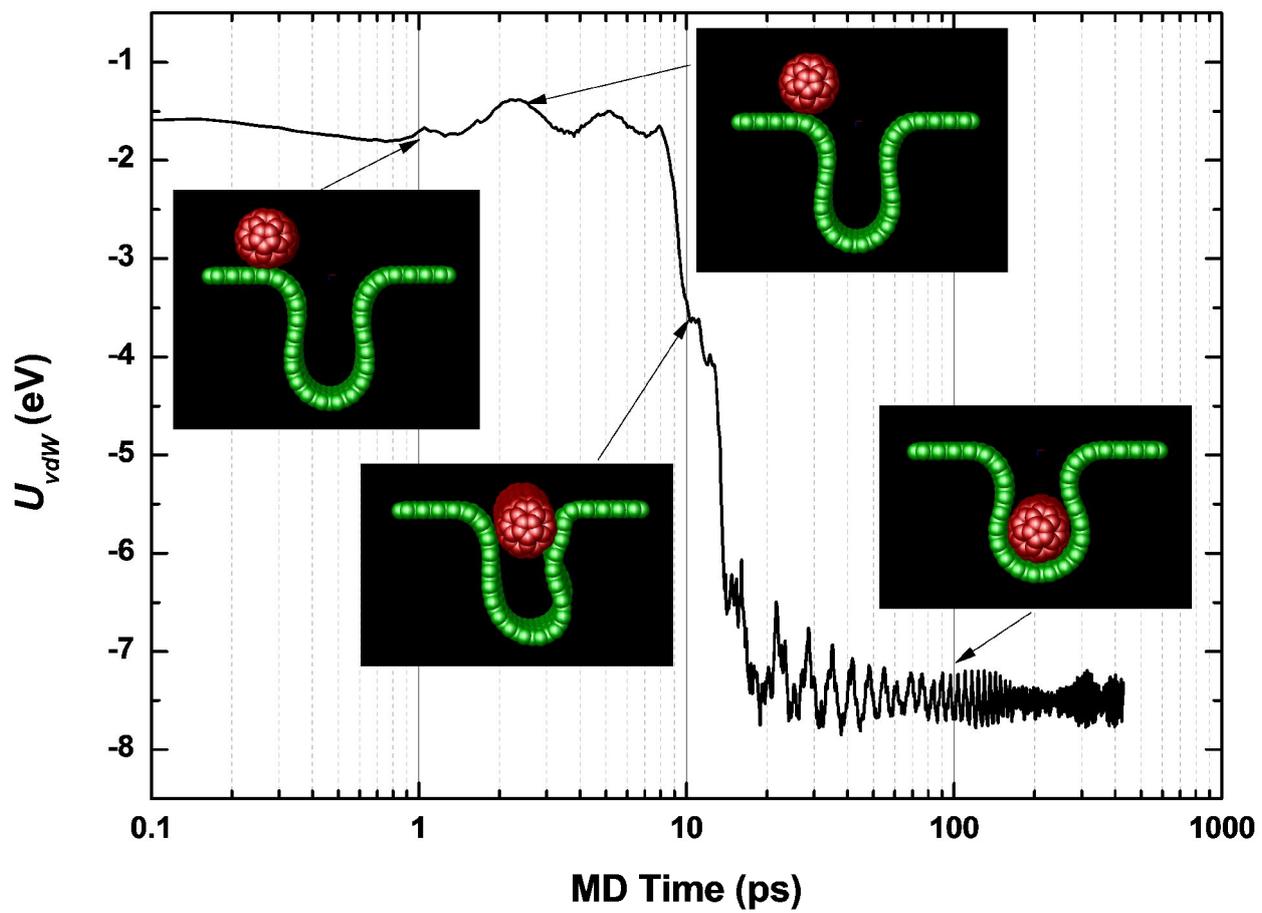

Figure 4a.

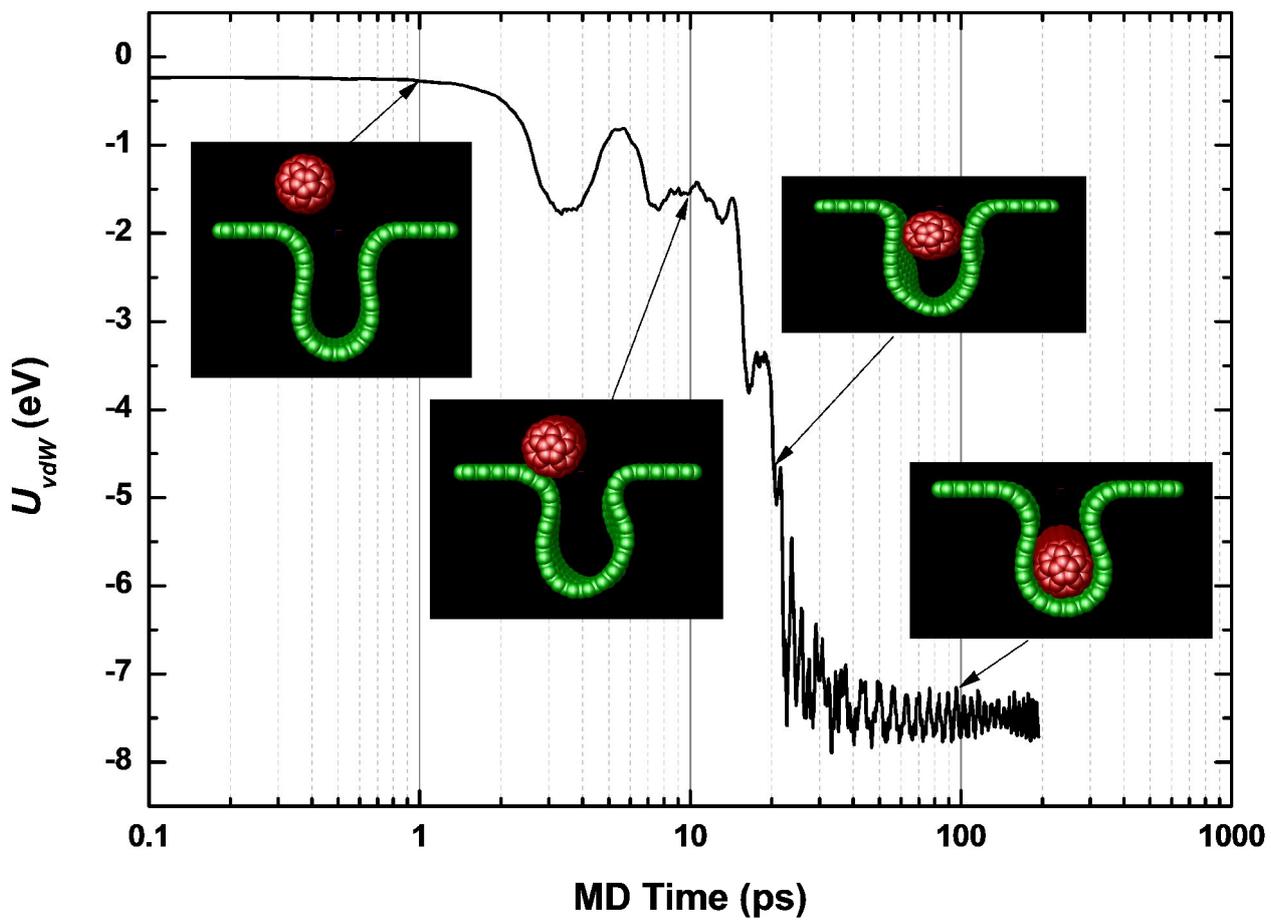

Figure 4b.

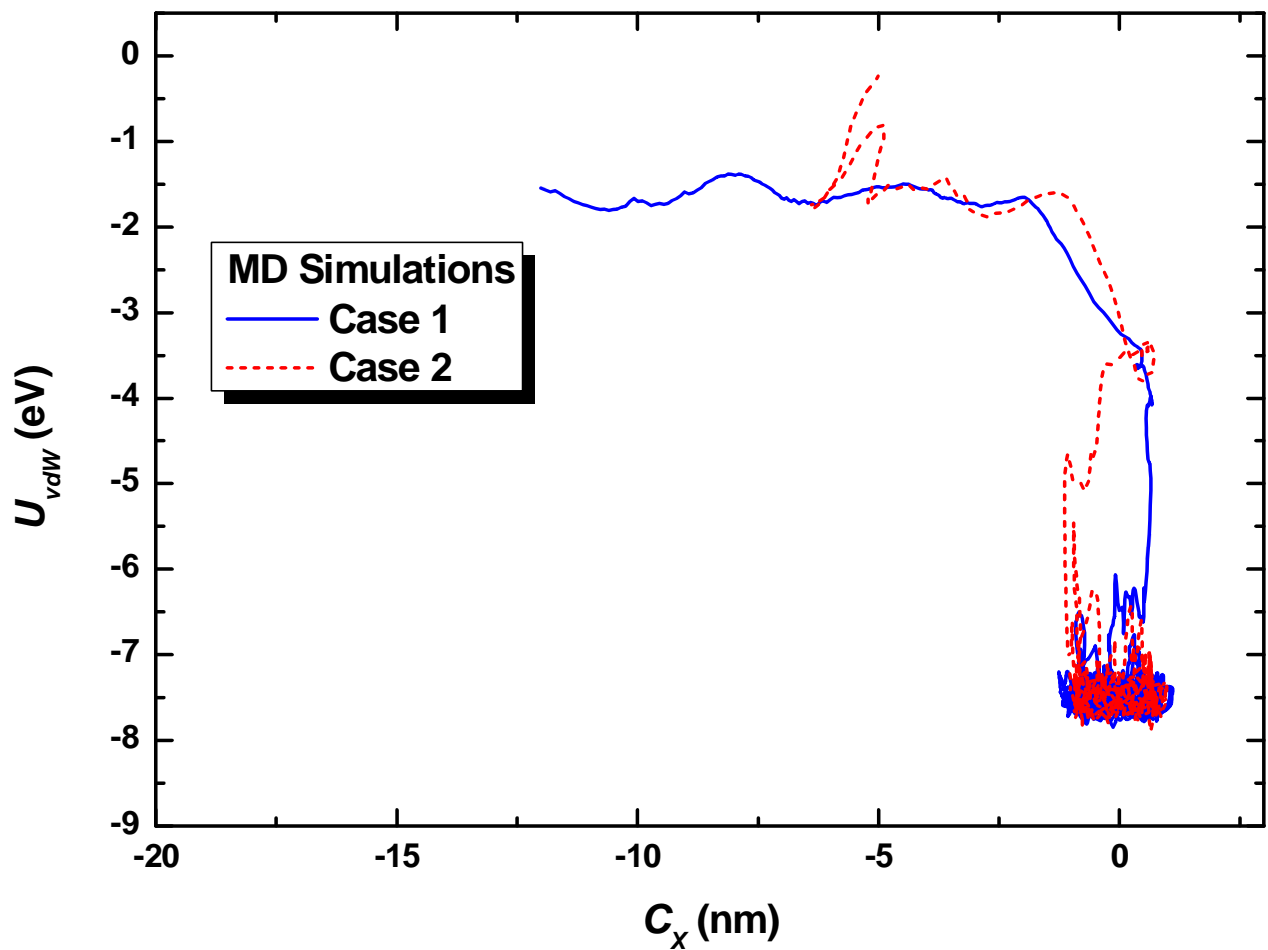

Figure 4c.

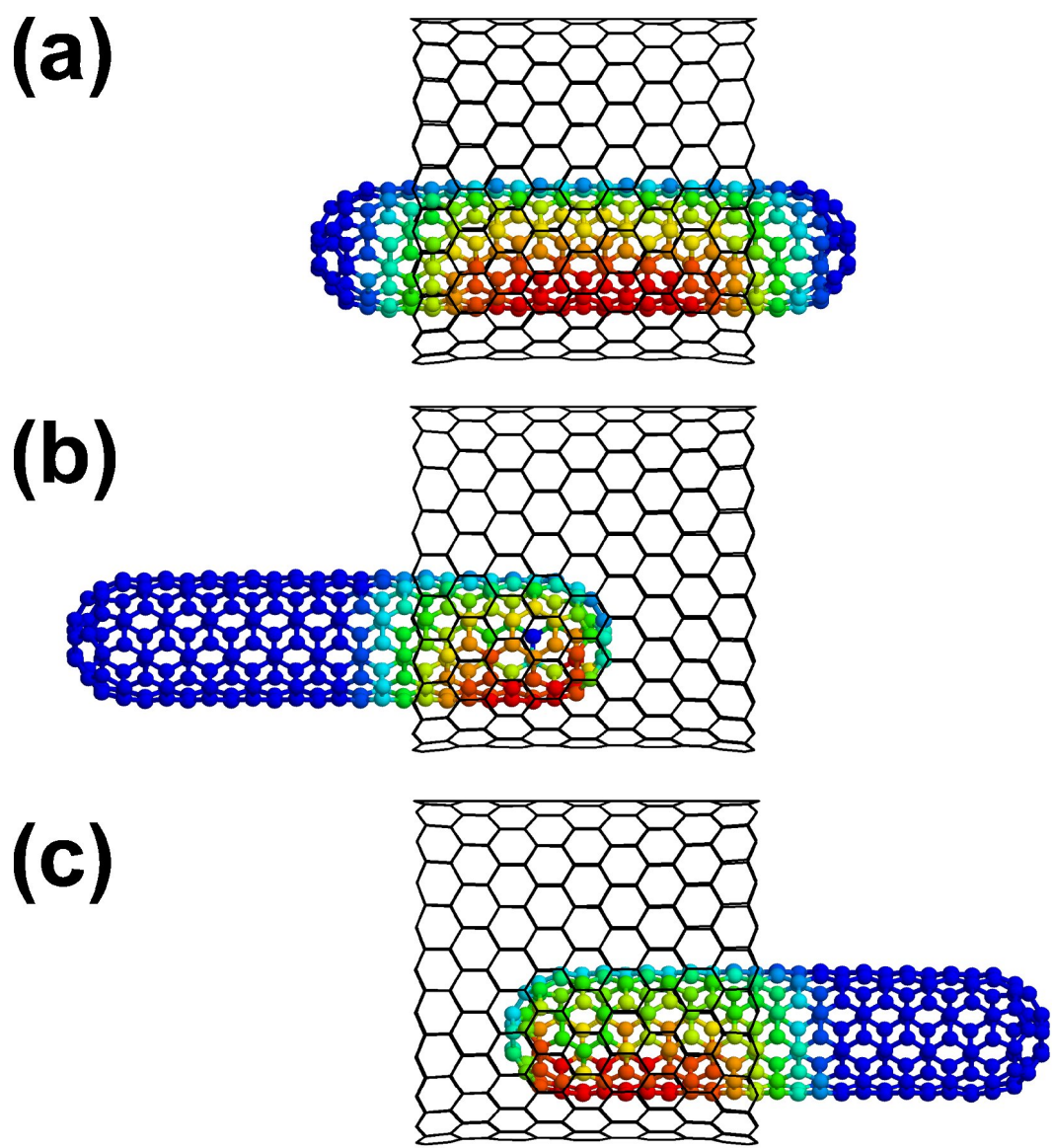

Figure 5.

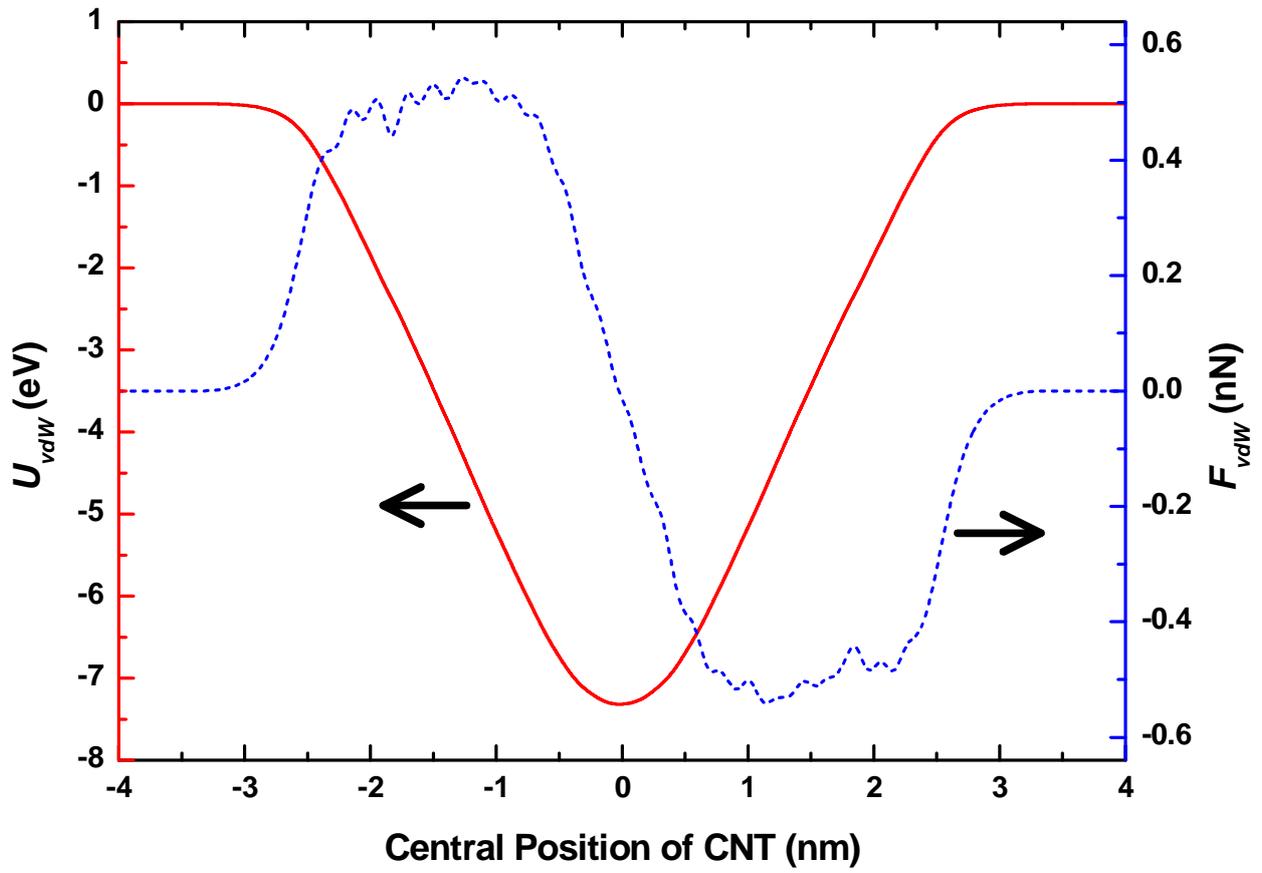

Figure. 6

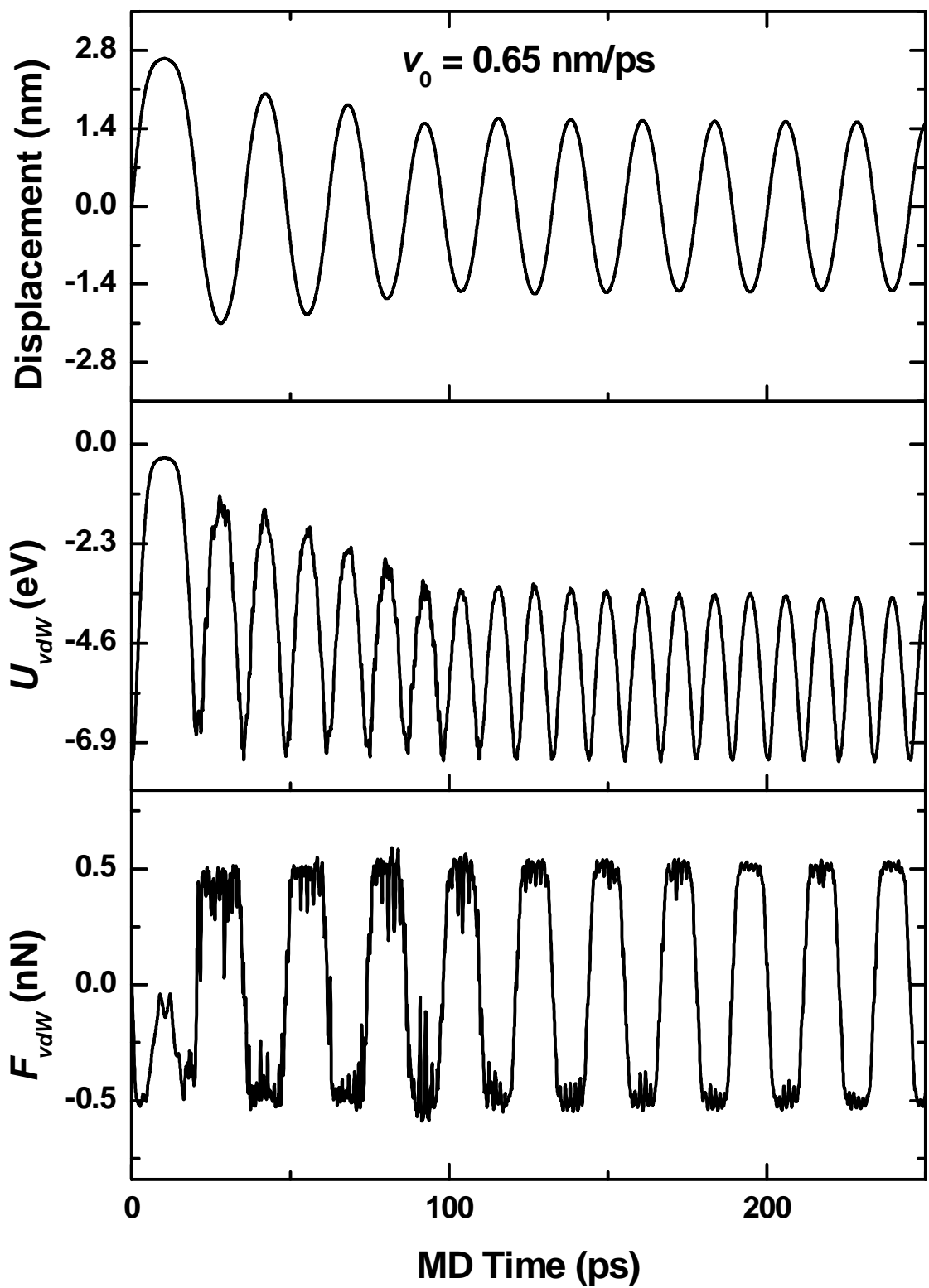

Figure 7.

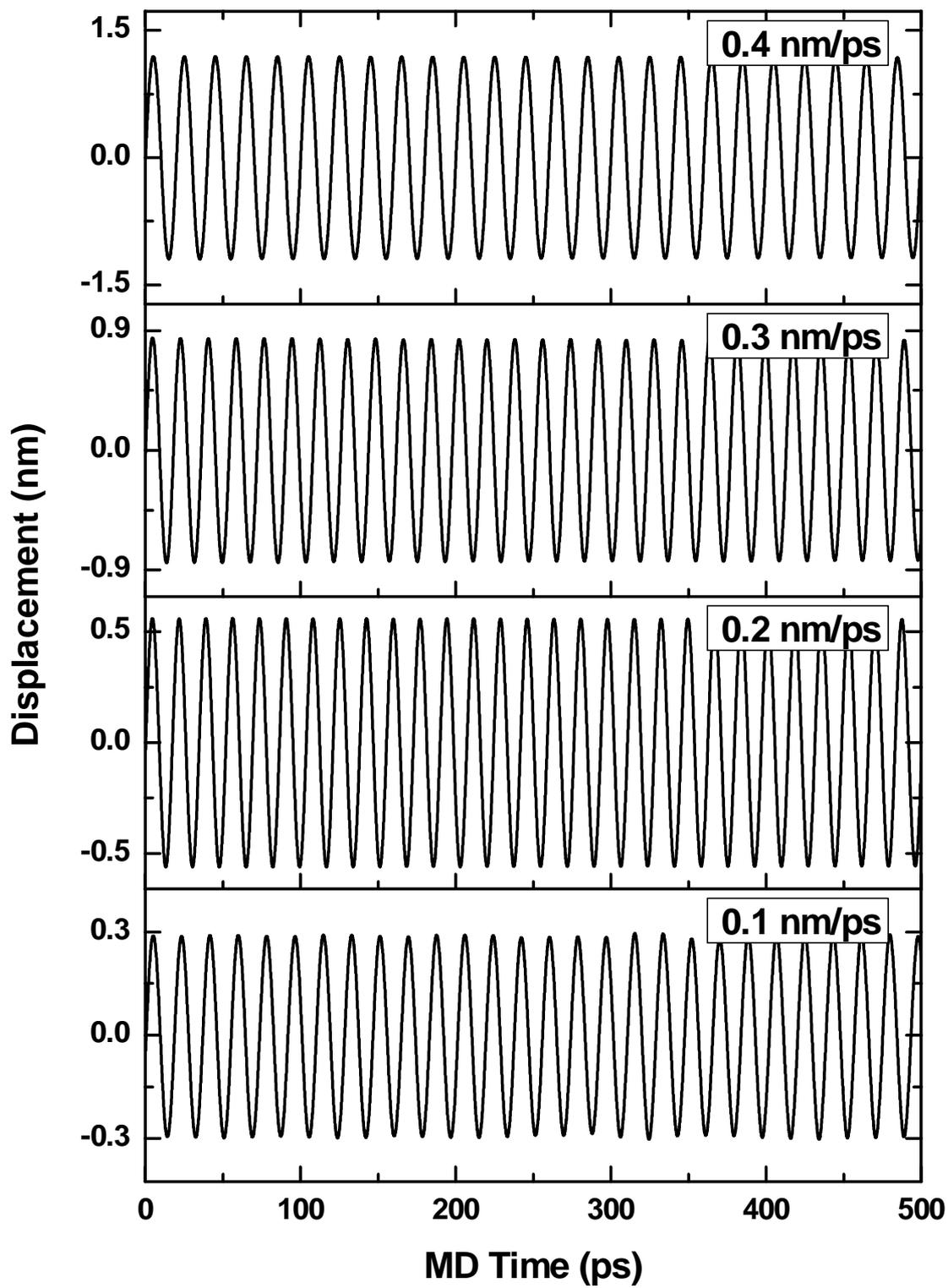

Figure 8a.

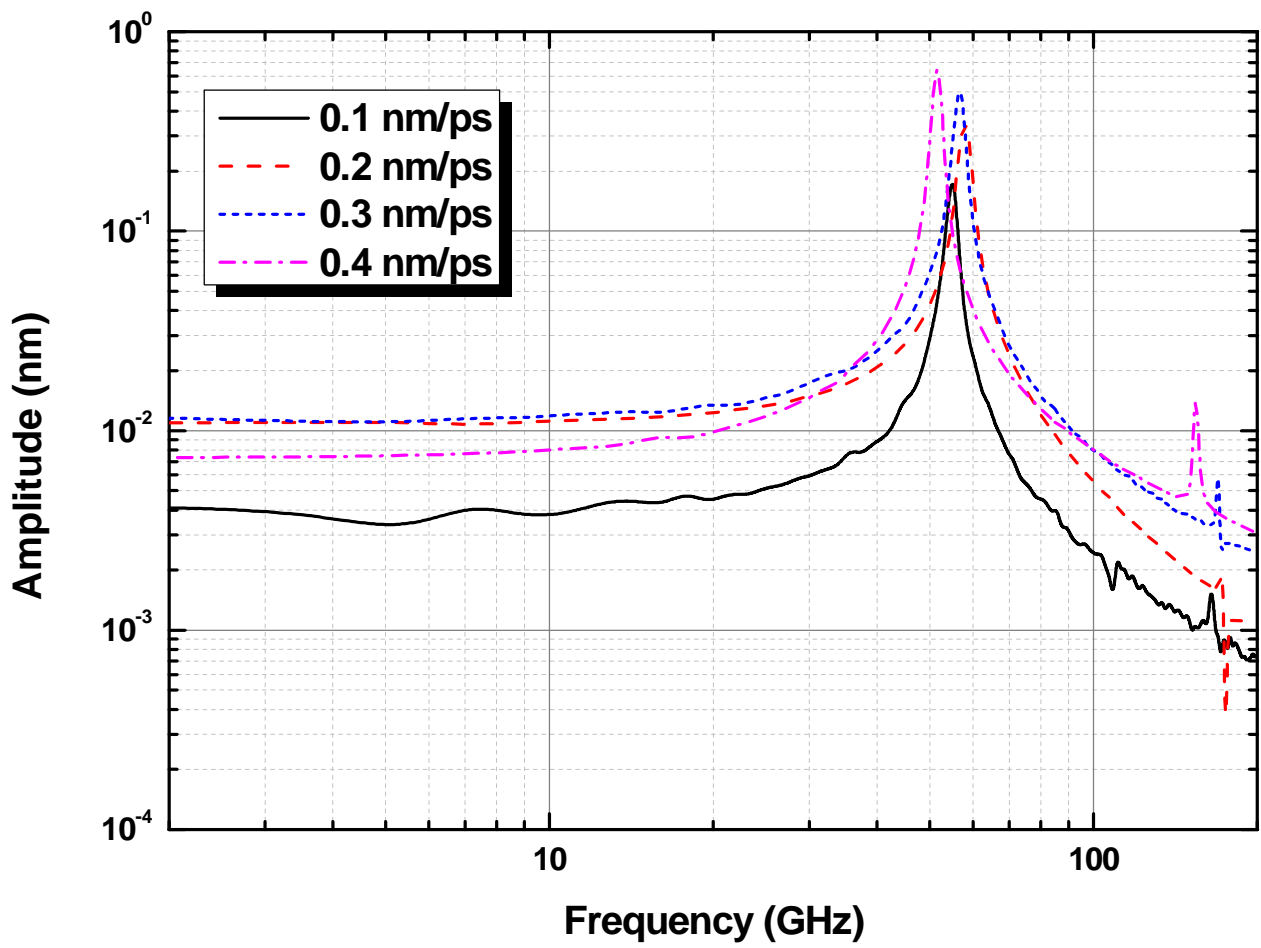

Figure 8b.

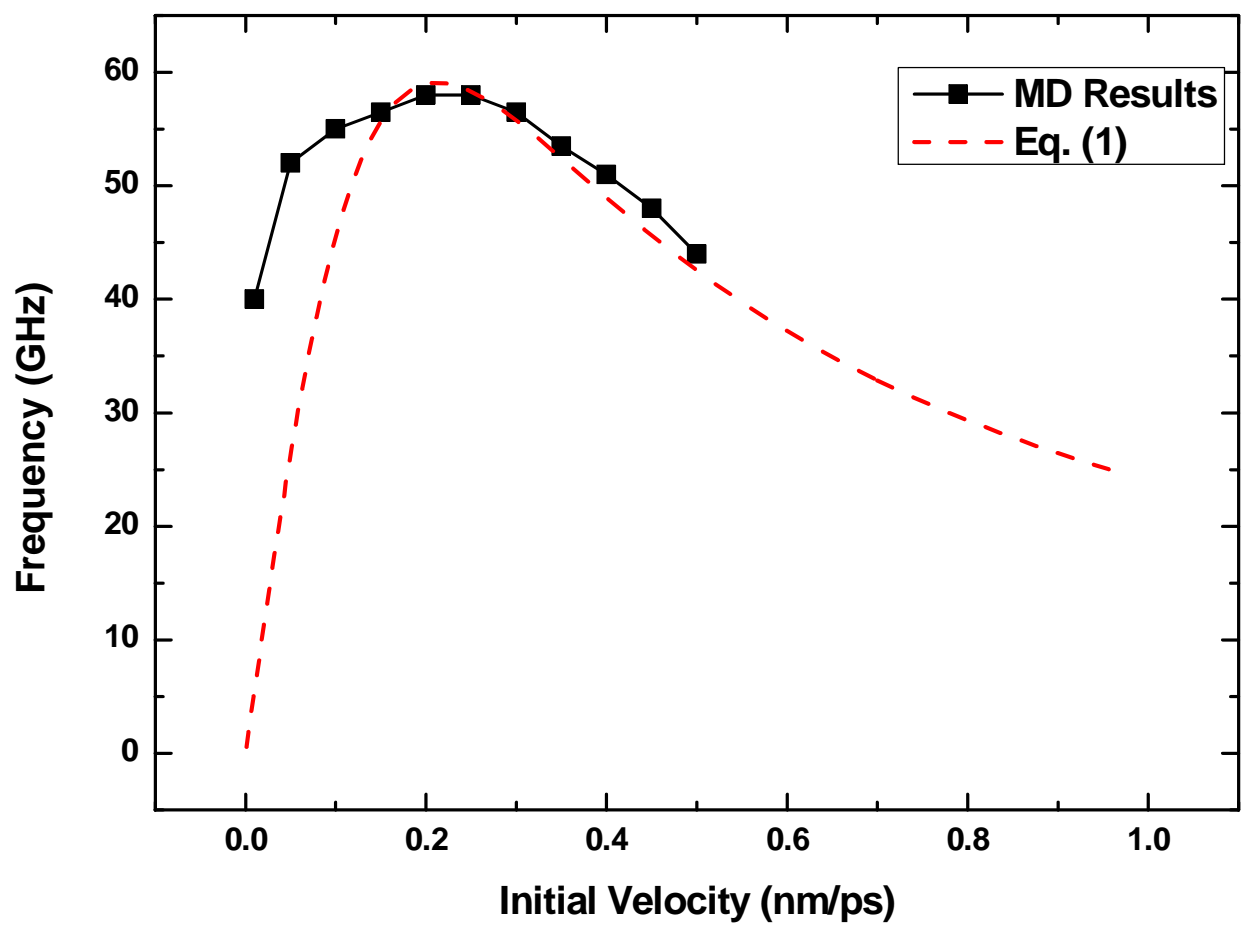

Figure 8c.